\documentclass[twocolumn,showkeys,showpacs,preprintnumbers,prd,superscriptaddress,nofootinbib,aps,10pt]{revtex4-1}
\usepackage{graphicx,epsf,bm,amsmath,amsfonts,amssymb,epstopdf,natbib,verbatim,multirow,appendix,array,diagbox,xcolor,tikz,hyperref,booktabs,siunitx}
\hypersetup{colorlinks=true,urlcolor=blue,citecolor=blue,linkcolor=blue,menucolor=blue,anchorcolor=blue,filecolor=blue}

\date{\today}

\definecolor{lime}{HTML}{A6CE39}
\DeclareRobustCommand{\orcidicon}{
\begin{tikzpicture}
\draw[lime, fill=lime] (0,0) 
circle [radius=0.16] 
node[white] {{\fontfamily{qag}\selectfont \tiny ID}};
\draw[white, fill=white] (-0.0625,0.095) 
circle [radius=0.007];
\end{tikzpicture}
\hspace{-3mm}
}

\foreach \x in {A, ..., Z}{\expandafter\xdef\csname orcid\x\endcsname{\noexpand\href{https://orcid.org/\csname orcidauthor\x\endcsname}{\noexpand\orcidicon}}}

\begin{document}

\title{Evaporating cosmologically coupled black holes}

\author{Marco Calz\`{a}\orcidA{}}
\email{marco.calza89@gmail.com}
\affiliation{Department of Physics, University of Trento, Via Sommarive 14, 38123 Povo (TN), Italy}
\affiliation{Trento Institute for Fundamental Physics and Applications (TIFPA)-INFN, Via Sommarive 14, 38123 Povo (TN), Italy}
\affiliation{Faculty of Sciences and Technology and CFisUC, University of Coimbra, Rua Larga, 3004-516 Coimbra, Portugal \looseness=-1}

\author{Davide Pedrotti\orcidB{}}
\email{davide.pedrotti-1@unitn.it}
\affiliation{Department of Physics, University of Trento, Via Sommarive 14, 38123 Povo (TN), Italy}
\affiliation{Trento Institute for Fundamental Physics and Applications (TIFPA)-INFN, Via Sommarive 14, 38123 Povo (TN), Italy}

\author{Massimiliano Rinaldi\orcidC{}}
\email{massimiliano.rinaldi@unitn.it}
\affiliation{Department of Physics, University of Trento, Via Sommarive 14, 38123 Povo (TN), Italy}
\affiliation{Trento Institute for Fundamental Physics and Applications (TIFPA)-INFN, Via Sommarive 14, 38123 Povo (TN), Italy}

\author{Sunny Vagnozzi\orcidD{}}
\email{sunny.vagnozzi@unitn.it}
\affiliation{Department of Physics, University of Trento, Via Sommarive 14, 38123 Povo (TN), Italy}
\affiliation{Trento Institute for Fundamental Physics and Applications (TIFPA)-INFN, Via Sommarive 14, 38123 Povo (TN), Italy}

\begin{abstract}
\noindent Cosmologically coupled black holes (CCBHs), whose masses evolve in response to the cosmological expansion, have recently attracted significant theoretical and observational interest. Existing studies have treated CCBHs as purely classical objects, neglecting the effect of Hawking radiation (HR), which competes with the cosmological coupling (CC) mechanism. We take a first step towards studying evaporating CCBHs, adopting a quasi-adiabatic approximation in which the HR rate is evaluated at the instantaneous CCBH mass, and modeling the CC mechanism through the phenomenological scaling of the mass with the scale factor, $M \propto a^k$. We show that, depending on the coupling strength $k$, even late-time CC activation can significantly delay Hawking evaporation, or lead to asymptotic CC-dominated mass growth, with important implications. We set limits on the abundance of primordial CCBHs from $\gamma$-ray observations, finding limits which are weaker than their uncoupled counterparts, as CCBHs are kept farther from the endpoint of evaporation for a longer time. Unlike standard primordial black holes, the CCBH formation and present-day masses no longer approximately coincide, even for formation masses $M_{\text{form}} \gtrsim 10^{15}\,{\text{g}}$. Therefore, the same population of primordial CCBHs may be subject to evaporation limits through its past emission history, as well as to other limits (such as microlensing) through its present-day mass.
\end{abstract}
\maketitle

\section{Introduction}
\label{sec:introduction}

Black holes (BHs) are usually modelled as being asymptotically flat and decoupled from the cosmological expansion~\cite{Frolov:1998wf,Fabbri:2005mw}, a description which is so far in excellent agreement with a host of BH-related observations across the gravitational and electromagnetic spectra~\cite{LIGOScientific:2016lio,Held:2019xde,Bambi:2019tjh,Vagnozzi:2019apd,Zhu:2019ura,Cunha:2019ikd,Banerjee:2019nnj,Banerjee:2019xds,Allahyari:2019jqz,Khodadi:2020jij,Kumar:2020yem,Gralla:2020srx,Volkel:2020xlc,Khodadi:2020gns,Pantig:2021zqe,EventHorizonTelescope:2021dqv,Khodadi:2021gbc,LIGOScientific:2021sio,Vagnozzi:2022moj,Uniyal:2022vdu,Pantig:2022ely,Ghosh:2022kit,Khodadi:2022pqh,KumarWalia:2022aop,Shaikh:2022ivr,Odintsov:2022umu,Oikonomou:2022tjm,Afrin:2022ztr,Pantig:2023yer,Gonzalez:2023rsd,Nozari:2023flq,daSilva:2023jxa,Uniyal:2023ahv,EventHorizonTelescope:2022xqj,Raza:2023vkn,Hoshimov:2023tlz,Chakhchi:2024tzo,Liu:2024lbi,Ditta:2024iky,Liu:2024lve,Khodadi:2024ubi,Nojiri:2024txy,Aktar:2024akk,Molla:2024zxi,Mustafa:2025cou,LIGOScientific:2025wao,Khodadi:2025upl,Nozari:2026wjo,Brax:2026cmh,Liang:2026rzy}. However successful, this description must ultimately be provisional at best, if anything because the Universe is not asymptotically flat at spatial infinity. While efforts to embed BHs within realistic boundary conditions describing an expanding Universe have been ongoing for nearly a century~\cite{McVittie:1933zz,Nolan:1998xs,Kaloper:2010ec,Lake:2011ni,daSilva:2012nh} (see also Ref.~\cite{Faraoni:2015ula} for a comprehensive review), only recently has it been appreciated that this may require BHs to be coupled to the cosmological expansion. Various unexpected dynamical phenomena emerge as a consequence, such as dynamical (comoving) horizons, or BH masses which, independently of accretion and mergers, grow with the scale factor of the Universe $a$. This cosmological coupling (CC) has strong theoretical support from at least two complementary arguments. First, within approaches which resolve the averaging ambiguity in the Friedmann-Lema\^{i}tre-Robertson-Walker (FLRW) equations, the stress-energy of relativistic compact objects can vary with the expansion rate \cite{Croker:2019mup}. Additionally, forcing an exactly static BH event horizon onto a time-dependent geometry results in a pathological naked singularity at the would-be horizon~\cite{Faraoni:2024ghi}; it has recently been argued that, because of this, BH horizons become comoving with an expanding Universe~\cite{Faraoni:2026jhk}.

On the observational side, the idea of cosmologically coupled BHs (CCBHs) has recently received significant interest. For instance, intriguing hints for CC-induced mass growth have been reported from observations of elliptical galaxies \cite{Farrah:2022ghw,Farrah:2023opk,Lacy:2023kbb,Farrah:2025ghw}, although these hints have been debated~\cite{Rodriguez:2023gaa,Andrae:2023wge,Gao:2023keg,Mlinar:2023fkk,Hayashi:2025frr}. The CC mechanism can also play a role for what concerns the unexpectedly large masses of certain binary BH systems as inferred from gravitational wave observations~\cite{Croker:2019kje,Amendola:2023ays}. Moreover, the CC mechanism has been shown to affect the amplitude of the astrophysical stochastic gravitational wave background from mergers of binary supermassive BHs in a detectable manner~\cite{Calza:2024qxn}, an effect which is highly relevant in light of the recent detection from pulsar timing arrays. Interestingly, for sufficiently strong mass growth, CCBHs can drive cosmic acceleration and thus be at the origin of dark energy (DE), potentially explaining the hints for dynamical DE inferred from recent Dark Energy Spectroscopic Instrument (DESI) data~\cite{Croker:2024jfg,Lei:2023mke}. Overall, while the question of the microscopic origin of CC is not yet settled, the CCBH framework is clearly one which is theoretically motivated and phenomenologically rich, explaining why the issue of CC (or absence thereof) is one which has attracted significant recent interest~\cite{Faraoni:2007es,Faraoni:2007gq,Faraoni:2008tx,Davidson:2012si,Croker:2019mup,Croker:2019kje,Croker:2020plg,Croker:2021duf,Cadoni:2022chn,Farrah:2022ghw,Farrah:2023opk,Mistele:2023fds,Cadoni:2023lum,Rodriguez:2023gaa,Parnovsky:2023wkc,Avelino:2023rac,Wang:2023aqe,Andrae:2023wge,Lei:2023mke,Sadeghi:2023cpd,Ghodla:2023iaz,Garcia-Bellido:2024tip,Yagdjian:2023yjf,Amendola:2023ays,Gao:2023keg,Gaur:2023hmk,Cadoni:2023lqe,Mlinar:2023fkk,Lacy:2023kbb,Dahal:2023hzo,Croker:2024jfg,Kovacik:2024xva,Cadoni:2024jxy,Lu:2024ppa,Faraoni:2024ghi,Calza:2024qxn,DES:2024ffp,Cadoni:2024rri,Farrah:2025ghw,Ahlen:2025owq,DESI:2025ffm,Lei:2025qff,Hayashi:2025frr,deLima:2025bqf,FelipeReis:2025qpf,Cadoni:2026ejk,Abilmazhinova:2026cog,Wu:2026xwo,Milos:2026ayt,Zhou:2026dti,Faraoni:2026jhk}.~\footnote{We note that the CC mechanism is most often discussed in connection with regular BHs, i.e.\ space-times which resolve or avoid the classical singularities present in the standard Schwarzschild and Kerr metrics, and which have been widely discussed in the recent literature (see e.g.\ Refs.~\cite{Bardeen:1968ghw,Hayward:2005gi,AyonBeato:1998ub,AyonBeato:1999rg,Easson:2002tg,Bronnikov:2005gm,Berej:2006cc,Bronnikov:2012ch,Rinaldi:2012vy,Culetu:2013fsa,Balart:2014cga,Culetu:2014lca,Ghosh:2014pba,Stuchlik:2014qja,Schee:2015nua,Johannsen:2015pca,Dymnikova:2015yma,Myrzakulov:2015kda,Fan:2016hvf,Sebastiani:2016ras,Toshmatov:2017zpr,Chinaglia:2017uqd,Frolov:2017dwy,Pacheco:2018mvs,Simpson:2019mud,Bertipagani:2020awe,Nashed:2021pah,Simpson:2021dyo,Franzin:2022iai,Chataignier:2022yic,Ghosh:2022gka,Lewandowski:2022zce,deFreitasPacheco:2023hpb,Boshkayev:2023rhr,Luongo:2023jyz,Luongo:2023aib,Giambo:2023zmy,Bonanno:2023rzk,Luongo:2023xaw,Sajadi:2023ybm,Javed:2024wbc,Ditta:2024jrv,Ovgun:2024zmt,Corona:2024gth,Konoplya:2024hfg,Pedrotti:2024znu,Bronnikov:2024izh,Kurmanov:2024hpn,Ovalle:2024wtv,Bolokhov:2024sdy,Agrawal:2024wwt,Banerjee:2024sao,Belfiglio:2024wel,Zhang:2024khj,Khodadi:2024efq,Estrada:2024uuu,KumarWalia:2024yxn,Li:2024ctu,Davies:2024ysj,Estrada:2024moz,Benavides-Gallego:2024hck,Frolov:2024hhe,Balart:2024rtj,Zhang:2024ney,Vertogradov:2025snh,Sajadi:2025prp,Xiong:2025hjn,Prado-Fuentes:2025nvl,Casadio:2025pun,Dialektopoulos:2025mfz,Harada:2025cwd,Fauzi:2025ldu,Kala:2025xnb,Capozziello:2025ycu,Pedrotti:2025idg,Urmanov:2025nou,Alonso-Bardaji:2025qft,Pinto:2025loq,Calza:2025mrt,Trivedi:2025vry,Neves:2025uoi,Capozziello:2025wwl,Asmanoglu:2025agc,Eichhorn:2025pgy,Ovalle:2025pue,Bonanno:2025dry,Jusufi:2025qgd,Khodadi:2025icd,Trivedi:2025agk,Zare:2025aek,Loc:2025mzc,Calza:2025yfm,Donmez:2025mcn,Battista:2026nsx,Arbelaez:2026eaz,Karakasis:2026wes,Wang:2026sqr,Arbey:2026koc,Skvortsova:2026ryl,Tello-Ortiz:2026dzd}). The CC framework can therefore potentially address two limitations of the standard BH description, i.e.\ asymptotically flat boundary conditions and classical singularities.}

While all existing studies of CCBHs treat them as classical objects, this cannot be the end of the story, as BHs are not purely classical. Indeed, semiclassical considerations concerning quantum fields evolving on a classical BH background indicate that BHs emit quasi-thermal radiation~\cite{Hawking:1974rv}: this is referred to as Hawking radiation (HR), and gradually decreases the BH mass. For stellar-mass and supermassive BHs, the evaporation timescale is significantly longer than the age of the Universe, justifying the fact that HR is neglected when studying astrophysical CCBHs. For sufficiently light BHs, whose evaporation timescale is shorter, the situation can be very different. This is particularly important in the case of primordial BHs (PBHs), which can play the role of dark matter (DM): indeed, some of the strongest limits on their abundance come from the absence of evaporation products, indicating that PBHs lighter than $\approx 10^{17}\,{\text{g}}$ cannot make up the entirety of the DM in the Universe~\cite{Khlopov:2008qy,Carr:2020xqk,Green:2020jor,Choudhury:2024aji,Shankaranarayanan:2026hnn}. If CC is a feature of BHs embedded in an expanding Universe, there is no a priori reason why the mechanism should only apply to astrophysical BHs. A very natural question is thus: \textit{what happens when Hawking evaporation and cosmological coupling are both at play}? Our work takes a first step towards addressing this question.

It is clear that the two effects act in opposite directions. As time moves on and the Universe expands, the CC mechanism increases the BH mass, while HR acts to decrease it. Therefore, one could expect the CC mechanism to potentially stabilize sufficiently heavy BHs against HR (or at least increase their lifetime), or conversely HR to eventually take over CC for sufficiently light BHs. As we shall show, the CC and HR contributions are intrinsically dynamical, i.e.\ the two relevant rates are time-dependent: a first-principles treatment would therefore require a fully dynamical space-time describing an evaporating cosmologically coupled BH. Such a metric is generically not available, even in the ``simpler'' case where only CC is present, which is why existing studies of CCBHs have mostly proceeded via phenomenological mass evolution ans\"{a}tze, for instance the often adopted $M(a) \propto a^k$~\cite{Croker:2019mup,Croker:2020plg,Croker:2021duf}. With these considerations in mind, as a zeroth order step, we find it natural to tackle the problem via a \textit{quasi-adiabatic approximation}: specifically, we evolve the BH mass under the combined effects of CC (via the usual phenomenological mass evolution ansatz) and HR, with the instantaneous evaporation rate taken to be the standard Hawking one evaluated at the corresponding value of the evolving mass. As we shall argue later, this approximation does not require a hierarchy between the CC and HR rates, but rather that the mass evolution be slow compared to the BH dynamical timescale: if that is the case, we can treat the BH as evolving through a sequence of quasi-static configurations. This condition can be met even in the phenomenologically interesting case where the CC and HR rates are comparable, and is expected to fail close to the endpoint of evaporation, where the semiclassical approximation is nevertheless known to be inadequate. In this work we expand on the above line of reasoning, studying the evolution of evaporating CCBHs within the quasi-adiabatic approximation.~\footnote{The possibility that the CC mechanism may counteract HR has already been noted earlier in Ref.~\cite{Ghodla:2023iaz}, where the authors estimated the minimum mass of PBHs that are stabilized in this way. Our analysis differs substantially from this earlier one in that we solve the full late-time joint evolution of HR and CC, and derive the resulting evaporation limits on primordial CCBHs.} We show that, under the simple phenomenological mass evolution ansatz $M \propto a^k$, the competition between CC and HR asymptotically leads to one of the two taking over: even when the asymptotic outcome is HR-dominated, however, the CC mechanism can significantly increase the lifetime of an evaporating BH. As an illustrative application, we compute evaporation constraints on light, cosmologically coupled PBHs, showing how the CC mechanism weakens these constraints relative to the uncoupled case, as it keeps BHs far from the (runaway) endpoint of the evaporation regime for a longer time.

The rest of this work is then organized as follows. In Sec.~\ref{sec:theory} we review two key theoretical ingredients of our study: Hawking radiation (Sec.~\ref{subsec:hr}) and the cosmological coupling mechanism (Sec.~\ref{subsec:ccbh}). In Sec.~\ref{sec:evaporatingccbh} we discuss our treatment of evaporating cosmologically coupled BHs, including the quasi-adiabatic approximation and its validity, as well as our methodology for setting constraints on the abundance of primordial cosmologically coupled BHs. Our results are presented and critically discussed in Sec.~\ref {sec:discussion}. Finally, in Sec.~\ref{sec:conclusions} we draw concluding remarks. Some useful analytical expressions relevant to our work, including a closed-form analytical solution for the mass evolution of evaporating CCBHs, are provided in Appendix~\ref{app:analytical}. Unless otherwise specified, we adopt units where $G=c=\hbar=1$.

\section{Theory}
\label{sec:theory}

Here we provide a brief review of the physics of Hawking radiation and the cosmological coupling mechanism.

\subsection{Hawking radiation}
\label{subsec:hr}

The existence of Hawking radiation is one of the most important predictions of quantum field theory in curved space-time, and originates from the fact that on a non-flat, time-dependent background, there is no unique notion of vacuum state. For a BH forming as the result gravitational collapse, the vacuum state in the asymptotic past is therefore seen by observers at future infinity as containing outgoing particles. In what follows we specialize to Schwarzschild BHs. In this case, the emission spectrum is approximately thermal, with a Hawking temperature given by the following~\cite{Hawking:1974rv}:
\begin{equation}
T_H=\frac{1}{8\pi M}\,,
\label{eq:th}
\end{equation}
recalling that we are working in natural units. In general, an evaporating BH will emit not only photons, but the whole spectrum of (kinematically accessible) particles of the underlying theory, with the energy of the emitted particles coming at the expense of the BH mass. An isolated, evaporating BH will therefore lose mass in time. The exact mass loss rate depends mostly on the spectrum of particles which are light enough to be emitted at a given BH temperature (or equivalently, through Eq.~(\ref{eq:th}), BH mass). The mass loss rate also depends on the so-called graybody factors (GBFs), which govern the deviation of the emitted spectrum from that of a blackbody due to the near-horizon geometry. All these effects are captured in full generality in the so-called (temperature-dependent) depletion functions $f_s(T)$.

For the purposes of our work, it is sufficient to follow Page's simpler treatment of the emission of massless fields by an evaporating Schwarzschild BH~\cite{Page:1976df,Page:1976ki,Page:1977um}. Specifically, we consider 3 light neutrino species and their corresponding antiparticles, one light charged lepton species (electrons and positrons), the photon, and the graviton, while neglecting heavier Standard Model particles. This approximation is well justified for the mass range relevant to our analysis. For sufficiently light BHs which are close to the evaporation threshold at the present time, species heavier than the electron are only emitted in the very final stages of the (runaway) evaporation process, during a time which is a tiny fraction of the total BH lifetime. This approximation is, of course, not meant to reproduce the endpoint of the evaporation process in full detail. Nevertheless, it does so at a level that is more than sufficient to capture the mass evolution due to HR, as well as the competition between HR and CC, which is the main focus of this work.

With these considerations in mind, the approximate Page coefficients (which already account for the relevant spin and particle-antiparticle multiplicity factors) can be expressed in terms of the geometrical-optics cross-section $\sigma_0=27\pi M^2$ as follows~\cite{Page:1976df}:
\begin{align}
&\sigma_\nu=\sigma_e \simeq 0.66852\,\sigma_0\,,
\label{eq:sigmanue} \\
&\sigma_\gamma \simeq 0.24044\,\sigma_0\,,
\label{eq:sigmagamma} \\
&\sigma_g \simeq 0.02748\,\sigma_0\,.
\label{eq:sigmag}
\end{align}
We now define the Hawking evaporation coefficient ${\cal C}_{\text{HR}}$ as follows:
\begin{equation}
{\cal C}_{\text{HR}} \equiv \frac{9M_P^3}{20480\pi\sigma_0t_P} \left [ \frac{7}{8}\left ( 3\sigma_\nu+\sigma_e \right )+\sigma_\gamma+\sigma_g \right ] \,,
\label{eq:fhr}
\end{equation}
with the factor $7/8$ accounting for the fermionic nature of neutrinos and electrons, and $M_P$ and $t_P$ being the Planck mass and Planck time respectively. With this definition, the evolution of the mass of a BH subject only to HR is controlled by the following equation:
\begin{equation}
\frac{dM}{dt} = \Gamma_{\text{HR}} \equiv -\frac{{\cal C}_{\text{HR}}}{M^2}\,.
\label{eq:dmdthr}
\end{equation}
where $\Gamma_{\text{HR}}$ is the HR rate. With these approximations, Eq.~(\ref{eq:dmdthr}) can be solved analytically. Considering a BH of mass $M_{\text{form}}$ at time $t_{\text{form}}$, the BH mass evolution at subsequent times is given by the following:
\begin{equation}
M(t)= \left [ M_{\text{form}}^3-3{\cal C}_{\text{HR}}(t-t_{\text{form}}) \right ] ^{1/3}\,,
\label{eq:mhrt}
\end{equation}
which holds until the endpoint of evaporation, when the quantity inside the square brackets vanishes. This occurs at an evaporation time $t_{\text{ev}}$ given by the following:
\begin{equation}
t_{\text{ev}}=t_{\text{form}}+\frac{M_{\text{form}}^3}{3{\cal C}_{\text{HR}}}\,,
\label{eq:tev}
\end{equation}
with $t_{\text{ev}}-t_{\text{form}} \propto M_{\text{form}}^3$ being the BH lifetime.

\subsection{Cosmologically coupled black holes}
\label{subsec:ccbh}

As discussed earlier, the idea of CC of BHs is not only theoretically very well-motivated, but also leads to concrete observational predictions. Lacking a well-established, agreed upon explicit solution for CCBHs, here we adopt a phenomenological parametrization of the effects of the CC mechanism. Following the seminal work of Refs.~\cite{Croker:2021duf,Farrah:2023opk}, we assume that the most relevant effect of CC is to induce a purely cosmological (i.e.\ independent of accretion, mergers, and other astrophysical events) growth of BH masses $M$, whose evolution with redshift is parametrized as follows:
\begin{equation}
M(z)=M\Theta(z-z_{\text{CC}})+M \left ( \frac{1+z_{\text{CC}}}{1+z} \right ) ^k \Theta(z_{\text{CC}}-z)\,,
\label{eq:mccz}
\end{equation}
where $\Theta$ denotes the usual Heaviside step function. The above parametrization therefore describes a (non-evaporating) BH whose mass is $M$ up to redshift $z_{\text{CC}}$, after which the CC mechanism turns on, and the mass grows with the scale factor $a=(1+z)^{-1}$ as $M \propto a^k$. In what follows, we shall refer to $k$ as the ``coupling strength''. The requirement that all causal observers perceive causal flux constrains this parameter to lie in the range $-3 \leq k \leq 3$, with $k=3$ being the maximum allowed value for causal material with positive energy density~\cite{Croker:2019mup}. However, we restrict our attention to the range $0 \leq k \leq 3$. Negative values of $k$ are not of interest to us as they would lead to mass depletion, which goes in the same direction of HR rather than competing with it.

Once the CC mechanism is turned on, the energy density of a population of CCBHs evolves as $\rho \propto a^{k-3}$, with the $a^{-3}$ factor coming from the evolution of the volume of the Universe with the scale factor. Comparing this to the evolution of the energy density of a fluid with equation of state $w$, $\rho \propto a^{-3(1+w)}$, we see that we can identify $w=-k/3$. This agrees with the well-known result of $k=3$ corresponding to a component behaving exactly as a cosmological constant, with $\rho={\text{const}}$ and $w=-1$. Unsurprisingly, this is in fact the value predicted by a population of regular BHs with vacuum interiors, i.e.\ de Sitter cores, whose masses grows in such a way as to exactly compensate the would-be dilution of the energy density due to the expansion of the Universe~\cite{Croker:2019mup,Farrah:2023opk}. On the other hand, $k=0$ corresponds to the uncoupled case where BHs behave as a matter component whose energy density redshifts as $a^{-3}$. Finally, the $k=1$ and $k=2$ cases lead to the CCBH energy density diluting as $\rho \propto a^{-2}$ and $\propto a^{-1}$ respectively, analogously to a population of cosmic strings ($k=1$) or domain walls ($k=2$). In the $k=1$ case, the evolution of the energy density is also identical to the evolution of the effective energy density associated to spatial curvature.

Since we later aim to combine HR and CC, it is useful to express the ``CC rate'', i.e.\ the rate of growth of the BH mass in time, due to the CC mechanism. This is given by the following:
\begin{equation}
\Gamma_{\text{CC}}=\frac{dM}{dt}=\frac{dM}{dz}\frac{dz}{dt}=-H(z)(1+z)\frac{dM}{dz}\,,
\label{eq:dmdtcc}
\end{equation}
where we have used the well-known time-redshift relation within a spatially flat FLRW Universe:
\begin{equation}
t(z) = \int_{z}^{\infty}\frac{dz'}{H(z')(1+z')}\,.
\label{eq:tz}
\end{equation}
Combining Eqs.~(\ref{eq:mccz},\ref{eq:dmdtcc}) we find that the CC rate is obviously $\Gamma_{\text{CC}}=0$ for $z>z_{\text{CC}}$. For $z<z_{\text{CC}}$, corresponding to times $t>t_{\text{CC}}$, carrying out the derivative leads to the following expression for the CC rate:
\begin{equation}
\Gamma_{\text{CC}}=kMH(z) \left ( \frac{1+z_{\text{CC}}}{1+z} \right ) ^k=kH(t)M(t)\,.
\label{eq:gammacc}
\end{equation}
The appearance of the factor of $H$, which arises from the Jacobian $dt/dz$, could have been guessed from first principles. Since CCBHs couple to the cosmological expansion, the ``natural'' timescale for this process is the Hubble time $1/H$, and correspondingly the ``natural'' rate is indeed (up to the coupling strength $k$) precisely $H$.

Two important remarks concerning the parametrization given in Eq.~(\ref{eq:mccz}) are in order before moving on. First, as in previous work~\cite{Calza:2024qxn}, we restrict our attention to values of the CC activation redshifts $z_{\text{CC}} \lesssim 3$. The main reason is that we are interested in the late-Universe phenomenology of the CC mechanism, particularly with regards to the possibility that CCBHs may play a role in cosmic acceleration. For cosmological parameter values close to the best-fit $\Lambda$CDM ones, $z_{\text{CC}} \sim 3$ corresponds to an activation time of $t_{\text{CC}} \sim 2.1\,{\text{Gyr}}$, which still leaves a long lever arm in cosmological time over which the CC mechanism can operate. From the practical point of view, this choice also allows us to neglect radiation in the background expansion rate.

One might also wonder why the CC mechanism should suddenly turn on at $z_{\text{CC}}$, rather than being active throughout the entire cosmological history, which is what one might expect if CC is a genuine property of compact objects embedded in an expanding Universe. Stated differently, one might wonder whether the sharp turn-on of the CC mechanism at redshift $z_{\text{CC}}$ should be taken literally. We do not intend this to be the case, but rather suggest viewing Eq.~(\ref{eq:mccz}) as a phenomenological late-time parametrization, with $z_{\text{CC}}$ denoting the epoch after which the scaling $M \propto a^k$ provides an adequate description of the mass growth. This distinction can be particularly important for PBHs, if these objects are cosmologically coupled. If the same mass scaling were extrapolated back to arbitrarily early times, the growth factor $((1+z_i)/(1+z))^k$ would be enormous given the extremely high formation redshifts $z_i$ associated with PBHs, and likely phenomenologically problematic. For instance, such a population of PBHs would dominate the energy density well before the epochs relevant for cosmic acceleration, and would be subject to extremely stringent early-Universe constraints, for instance Big Bang Nucleosynthesis, the Cosmic Microwave Background, and structure formation. This problem has already been stressed in Ref.~\cite{Ghodla:2023iaz}, where early-forming primordial CCBHs were argued to be observationally viable only if the CC mechanism turns on sufficiently late. In this work, we therefore treat $z_{\text{CC}}$ as the epoch after which the simple scaling given in Eq.~(\ref{eq:mccz}) is a valid late-time description, rather than a first-principles prediction. The CC mechanism may well be active for $z \gg z_{\text{CC}}$, but would require a drastically different scaling in order to be phenomenologically acceptable: a complete treatment of the early-time behaviour of the CC mechanism would therefore require a dedicated model, which goes beyond the simple ansatz of Eq.~(\ref{eq:mccz}), and lies well beyond the scope of this work. We stress that our parametrization, while deliberately phenomenological, is tailored to the late Universe and current observational signatures, and has been adopted in several other works~\cite{Farrah:2023opk,Calza:2024qxn}, making it easier to place our analysis in the context of existing literature.

\section{Evaporating cosmologically coupled black holes}
\label{sec:evaporatingccbh}

As anticipated earlier, treating CCBHs as purely classical objects cannot be the end of the story. With this in mind, we now combine the two ingredients discussed above, i.e.\ Hawking radiation and cosmological coupling, to study the evolution of evaporating CCBHs.

\subsection{Mass evolution}
\label{subsec:evolution}

We recall that for an ordinary (non-cosmologically coupled) BH, the mass loss due to Hawking evaporation is governed by the rate given in Eq.~(\ref{eq:dmdthr}), whereas the mass gain for a non-evaporating CCBH is governed by the rate given in Eq.~(\ref{eq:dmdtcc}), with both rates being manifestly time-dependent. The question then becomes how to combine the two ingredients. The main obstacle in this direction is the absence of a fully dynamical metric describing an evaporating CCBH. With this in mind, the natural zeroth order step is to adopt a \textit{quasi-adiabatic approximation}. Specifically, we assume that the total rate of change in the BH mass is given by the sum of the HR [Eq.~(\ref{eq:dmdthr})] and CC [Eq.~(\ref{eq:dmdtcc})] contributions, leading to the following mass evolution equation:
\begin{equation}
\frac{dM(t)}{dt}=\Gamma_{\text{HR}}[M(t)]+\Gamma_{\text{CC}}[M(t),t]\,,
\label{eq:dmdthrcc}
\end{equation}
where we have explicitly spelled out the time-dependence of the mass entering the two rates. We note that $\Gamma_{\text{HR}}$ carries an implicit dependence on time through the instantaneous BH mass, whereas $\Gamma_{\text{CC}}$ carries an explicit time dependence through its dependence on the cosmological background expansion rate $H(t)$. The above Eq.~(\ref{eq:dmdthrcc}) should be understood as a phenomenological ansatz, itself relying on three logically distinct assumptions. The first is of course the quasi-adiabatic approximation, whose regime of validity we discuss in much more detail later. The second is an ``instantaneous Schwarzschild approximation'', i.e.\ the assumption that the BH evolves through instantaneous Schwarzschild (as opposed to different space-times) configurations with time-dependent masses $M(t)$. Therefore, all relevant quantities such as the BH horizon radius, surface gravity, Hawking temperature, Bekenstein-Hawking entropy, and GBFs, are assumed to retain the same functional form as their standard Schwarzschild counterparts, subject only to the replacement $M \to M(t)$.~\footnote{Within the same instantaneous Schwarzschild approximation, we can also assign to the evaporating CCBH the standard Bekenstein-Hawking entropy evaluated at the instantaneous mass, i.e.\ $S_{\text{BH}}(t)=4\pi M^2(t)$. The HR and CC terms then give competing effects for what concerns the evolution of $S_{\text{BH}}$, with HR decreasing it, and the CC mechanism increasing it. However, we believe it would be premature to interpret this in terms of a generalized entropy formula. The latter requires specifying the microphysical sector responsible for the CC mechanism, which we remain agnostic about within our deliberately phenomenological treatment thereof: we therefore defer a complete first and second law analysis of evaporating CCBHs to future work.} The third assumption is the additivity of the HR and CC rates, itself a reasonable approximation in the absence of a fully dynamical metric describing an evaporating CCBH.

We now distinguish three relevant times during the cosmic history. The first is the formation time $t_{\text{form}}$, i.e.\ the time at which the CCBH forms. For primordial CCBHs, formation takes place extremely early, typically shortly after inflation, so effectively $t_{\text{form}} \to 0$. The second is the ``CC activation time'' $t_{\text{CC}}$ already introduced in Eq.~(\ref{eq:mccz}), i.e.\ the time at which our CC parametrization becomes active. Finally, conforming to standard cosmological notation, we denote by $t_0$ the present time. In our later study of evaporation constraints on primordial CCBHs, the hierarchy $t_{\text{CC}} \gg t_{\text{form}}$ holds, whereas within the mass range of interest early mass loss from HR turns out to be negligible: under these conditions, we can therefore safely set $t_{\text{form}} \to 0$, while keeping the notation $t_{\text{form}}$ when useful for clarity. At the three relevant times $t_{\text{form}}$, $t_{\text{CC}}$, and $t_0$, we denote the instantaneous CCBH mass by $M_{\text{form}}=M(t_{\text{form}})$, $M_{\text{CC}}=M(t_{\text{CC}})$, and $M_0=M(t_0)$. We can now write the mass evolution equation explicitly in the two regimes prior to and after CC activation. For $t<t_{\text{CC}}$, we have $\Gamma_{\text{CC}}=0$, so that Eq.~(\ref{eq:dmdthrcc}) reduces to the following:
\begin{equation}
\frac{dM(t)}{dt}=-\frac{{\cal C}_{\text{HR}}}{M^2(t)}\,,
\label{eq:dmdthrcctltcc}
\end{equation}
whereas for $t>t_{\text{CC}}$ the mass evolution equation is given by the following:
\begin{equation}
\frac{dM(t)}{dt}=-\frac{{\cal C}_{\text{HR}}}{M^2(t)}+kH(t)M(t)\,.
\label{eq:dmdthrcctgtcc}
\end{equation}
Once initial conditions for $M(t_{\text{form}})$ or $M(t_{\text{CC}})$ are set, Eq.~(\ref{eq:dmdthrcc}), or equivalently its piecewise form given by Eqs.~(\ref{eq:dmdthrcctltcc},\ref{eq:dmdthrcctgtcc}), can be solved to determine the CCBH mass at any time, including its present-day mass $M_0$. If the solution reaches $M(t)=0$ at some finite time $t=t_{\text{ev}}$, we identify $t_{\text{ev}}$ as the CCBH evaporation time. Correspondingly, the CCBH lifetime is then given by $t_{\text{ev}}-t_{\text{form}}$, which reduces to $\approx t_{\text{ev}}$ for primordial CCBHs. We note that in this regime $M_0$ is not defined.~\footnote{Throughout this work we adopt the standard semiclassical description of Hawking evaporation. However, this description is expected to be at best incomplete after the Page time, when the backreaction due to the emitted radiation becomes non-negligible and may slow down the evaporation process, requiring additional ingredients for a complete description. A well-studied example in this sense is the memory burden effect (see e.g.\ Refs.~\cite{Dvali:2020wft,Thoss:2024hsr,Kohri:2024qpd,Zantedeschi:2024ram,Chianese:2024rsn,Montefalcone:2025akm,Dvali:2025ktz,Chianese:2025wrk,Dondarini:2025ktz,Dvali:2026tia}). Accounting for such effects would require going beyond the semiclassical approximation, and is well beyond the scope of this work.}

To solve Eq.~(\ref{eq:dmdthrcc}) once initial conditions are assigned, the background expansion history $H(t)$ entering Eq.~(\ref{eq:dmdthrcctgtcc}) needs to be specified. In practice we start with a prescribed $H(z)$, numerically invert the standard FLRW time-redshift relation [Eq.~(\ref{eq:tz})] to obtain $z(t)$, and use this to obtain $H(t)=H(z(t))$. In what follows, we assume a spatially flat $\Lambda$CDM background, whose expansion rate is given by the following:
\begin{equation}
H(z) \simeq H_0\sqrt{\Omega_m(1+z)^3+(1-\Omega_m)}\,,
\label{eq:hzlcdm}
\end{equation}
where we take the present-day matter density parameter to be $\Omega_m=0.315$, close to the best-fit \textit{Planck} $\Lambda$CDM cosmology~\cite{Planck:2018vyg}. We note that the approximation in Eq.~(\ref{eq:hzlcdm}) is valid exclusively at late times ($z \ll 1000$), when the radiation component can be neglected, and therefore the present-day density parameter for the cosmological constant is $\Omega_{\Lambda} \approx 1-\Omega_m$. This approximation is appropriate for our study, since we are assuming that the CC mechanism switches on at late times, $z_{\text{CC}} \lesssim 3$.

Three comments are now in order. First, we stress that, while we have adopted this choice of $H(z)$ for simplicity, any expansion history can be inserted into Eq.~(\ref{eq:dmdthrcc}), with the corresponding $t(z)$ relation computed and inverted numerically. Next, although in Sec.~\ref{subsec:ccbh} we noted that a CCBH population with coupling $k$ is associated to an effective equation of state $w=-k/3$, here we are not assuming that the CCBHs are the main source of the background expansion (whose rate would have to depend on $k$), but rather treat $H(z)$ as an externally specified cosmological background. For the purposes of our phenomenological study, this is a suitable approximation, especially considering that the background expansion rate given by Eq.~(\ref{eq:hzlcdm}) is overall in good agreement with a host of precision cosmological data, in spite of two major elephants in the room, i.e.\ the Hubble tension~\cite{Verde:2019ivm,DiValentino:2020zio,DiValentino:2021izs,Perivolaropoulos:2021jda,Shah:2021onj,Abdalla:2022yfr,DiValentino:2022fjm,Hu:2023jqc,Vagnozzi:2023nrq,Verde:2023lmm,CosmoVerseNetwork:2025alb,Cai:2026swf,Schoneberg:2026eys,Schoneberg:2026vaf,Wang:2026jxd} and the recent DESI hints for dynamical DE~\cite{Cortes:2024lgw,Colgain:2024xqj,Carloni:2024zpl,Giare:2024gpk,Jiang:2024xnu,RoyChoudhury:2024wri,Giare:2025pzu,DESI:2025zgx,Colgain:2025nzf,DESI:2025wyn,RoyChoudhury:2025dhe,Chaudhary:2025pcc,RoyChoudhury:2025iis,Li:2025vuh,Capozziello:2025qmh,Li:2026asg,Giare:2026oti}. While there is also some disagreement regarding the value of $\Omega_m$~\cite{Sakr:2023hrl,Akarsu:2024qiq,Pedrotti:2024kpn,Colgain:2024mtg,Lynch:2025ine,Lee:2025hjw,Lee:2025kbn,Wang:2025znm,Weiner:2026sfm,Shlivko:2026jxa}, we note that the specific value thereof does not play a major role in our results. Finally, for the specific matter plus cosmological constant background given in Eq.~(\ref{eq:hzlcdm}), useful closed-form analytical expressions for $t(z)$, $z(t)$, and $M(t)$ are available. We have chosen not to show these expressions in the main text as they are of limited practical use, and especially do not generalize straightforwardly to expansion histories which deviate from the chosen one: nevertheless, these expressions are useful as an analytical check of our numerical implementation, and may be of interest in their own right, so we provide them in Appendix~\ref{app:analytical}.

Before moving on, a few considerations on our quasi-adiabatic approximation and its validity regime are in order. This approximation does not require a hierarchy between $\Gamma_{\text{HR}}$ and $\Gamma_{\text{CC}}$. Rather, the relevant hierarchy involves the BH dynamical timescale $t_{\text{dyn}}$, and the typical timescale over which the BH mass changes, which we denote by $t_M$. Temporarily restoring $c$ and $G$, the BH dynamical timescale can be estimated as follows:
\begin{equation}
t_{\text{dyn}} \approx \frac{2GM}{c^3} \,.
\label{eq:tdyn}
\end{equation}
This is of order the light-crossing timescale, and is the characteristic time over which the near-horizon geometry can ``adjust'' to changes in the BH parameters. Concerning the mass evolution timescale, a sensible estimate for this quantity is instead the inverse of the logarithmic rate of mass change, i.e.\ the following:
\begin{align}
t_M(t) &\equiv \left \vert \left ( \frac{d\ln M(t)}{dt} \right ) ^{-1} \right \vert = \left \vert \frac{M(t)}{\dfrac{dM(t)}{dt}} \right\vert \nonumber\\
&= \left \vert \frac{M(t)}{\Gamma_{\rm HR}[M(t)]+\Gamma_{\rm CC}[M(t),t]} \right \vert \,.
\label{eq:tm}
\end{align}
Clearly, both $t_{\text{dyn}}$ and $t_M$ are time-dependent quantities, with the time dependence inherited through the time dependence of $M(t)$. The condition under which our quasi-adiabatic approximation is valid is then the following:
\begin{equation}
t_{\text{dyn}}(t) \ll t_M(t)\,.
\label{eq:validity}
\end{equation}
If the above condition is met, the evaporating CCBH can effectively be pictured as evolving through a sequence of configurations which are, instant by instant, quasi-static. Stated differently, Eq.~(\ref{eq:validity}) sets the requirement that the change in BH mass is negligible during one BH dynamical time. Under this condition, the geometry evolves slowly enough that we can treat the BH, instant by instant, as a quasi-static Schwarzschild BH with mass $M(t)$, which justifies the use of the instantaneous Hawking rate for Schwarzschild BHs. We stress that Eq.~(\ref{eq:validity}) does not require a hierarchy between $\Gamma_{\text{HR}}$ and $\Gamma_{\text{CC}}$. Our quasi-adiabatic approximation therefore also applies in regimes where the two are potentially comparable, provided that under their combined effects, the BH mass changes slowly compared to $t_{\text{dyn}}$. For the range of parameters considered in our study, i.e.\ $10^{14}\,{\text{g}} \lesssim M_{\text{form}} \lesssim 10^{18}\,{\text{g}}$, Eq.~(\ref{eq:validity}) is safely satisfied throughout most of the evolution, since $t_{\text{dyn}} \ll 10^{-20}\,{\text{s}}$, enormously smaller than both the typical CC and HR timescales. Of course, it is expected to fail close to the endpoint of evaporation, when $t_M(t) \to 0$ as $t \to t_{\text{ev}}$. Nevertheless, we know that the semiclassical approximation is bound to break down in the final stages of the evaporation process (see also footnote~3). Therefore, the failure of the quasi-adiabatic condition close to the endpoint of evaporation is not a limitation of our treatment, but coincides with the regime where our semiclassical HR description is no longer expected to be reliable.

For what concerns departures from the instantaneous Schwarzschild approximation, we expect these to be controlled (and suppressed) mostly by two small parameters. The first is the ratio $t_{\text{dyn}}(t)/t_M(t)$, i.e.\ the same parameter which controls the validity of the quasi-adiabatic approximation in Eq.~(\ref{eq:validity}). The second is instead $H(t)t_{\text{dyn}}(t)\sim H(t)r_h(t)/c$, i.e.\ the ratio of the BH horizon to the Hubble scale. Again, for the range of parameters considered in this work, both quantities are extremely small, except close to the endpoint of evaporation, where the semiclassical description of HR is expected to break down in any case, as stressed earlier. We thus expect corrections to the instantaneous Schwarzschild approximation to be small in the regime of interest.

In closing, we note that the same quasi-adiabatic logic is actually already implicitly present in the standard treatment of HR, e.g.\ in Eq.~(\ref{eq:dmdthr}), which essentially computes the instantaneous mass loss rate on a stationary Schwarzschild background. The motivation for adopting this approximation is actually similar to ours: there is no known exact, realistic dynamical metric for four-dimensional evaporating BHs which is a solution to the semiclassical Einstein equations and takes into account the back-reaction of HR on the underlying space-time (see e.g.\ Ref.~\cite{Thiemann:2024nmy}).~\footnote{Some progress on the problem of backreaction has been recently achieved for acoustic analogs of BHs~\cite{Balbinot:2025upx}.} Our treatment in Eq.~(\ref{eq:dmdthrcc}) simply extends this approximation by including the additional contribution from the CC mechanism. In a similar vein, we note that the same quasi-adiabatic approximation is implicitly adopted in several other BH-related contexts, e.g.\ BH superradiance, where the backreaction of the superradiant cloud on the BH geometry is often neglected, and the BH is treated instant by instant as being a quasi-static Kerr BH with mass and angular momentum determined by the superradiant evolution (see for instance Refs.~\cite{Brito:2014wla,Brito:2015oca,Roy:2019esk,Creci:2020mfg,Khodadi:2021owg,Roy:2021uye,Chen:2022nbb,Khodadi:2022dyi,Chen:2022kzv,Jha:2022tdl}). All these considerations justify the use of the quasi-adiabatic approximation, and in particular of Eq.~(\ref{eq:dmdthrcc}), in the present work.

\subsection{Constraints on the abundance of primordial cosmologically coupled black holes}
\label{subsec:constraints}

We now briefly outline the methodology utilized to set constraints on the abundance of primordial CCBHs, from the absence of their evaporation products. We follow the same methodology adopted by some of us to set evaporation constraints on (non-cosmologically coupled) primordial regular BHs in Refs.~\cite{Calza:2024fzo,Calza:2024xdh,Calza:2025mwn}, and encourage the reader to consult these works for further technical details.

We begin by computing the primary photon spectrum emitted by an evaporating CCBH whose instantaneous mass is $M$, i.e.\ the number of photons emitted per unit time per unit energy $E_{\gamma}$:
\begin{equation}
\Phi_{\gamma}(E_{\gamma},M)=\frac{1}{\pi}\sum_{l}\frac{(2\ell+1)\Gamma^{s=1}_{l}(E_{\gamma},M)}{e^{E_{\gamma}/T_H(M)}-1}\,,
\label{eq:phigamma}
\end{equation}
where $T_H(M)=1/8\pi M$ is the Hawking temperature, $\Gamma^{s=1}_{l}$ are the GBFs for spin-$1$ perturbations propagating on a Schwarzschild background, and the sum runs over the angular momentum quantum number $\ell$. Since we are implicitly considering a spherically symmetric background (as the impact of rotation on the CC mechanism remains yet to be studied), modes of different azimuthal quantum number $m$ at fixed $\ell$ are degenerate. This explains the appearance of the $(2\ell+1)$ factor in the numerator of Eq.~(\ref{eq:phigamma}). We compute the GBFs using the publicly available \texttt{GrayHawk} code~\cite{Calza:2025whq,Calza:2026wuf}, developed by one of us. It is important to note that the spectrum in Eq.~(\ref{eq:phigamma}) is time-dependent, as it inherits the time dependence of $M(z)$, due to HR and CC, in both the GBFs and temperature. Let us recall that, for Schwarzschild BHs, the GBFs are universal functions of the (dimensionless in appropriate units) combination $E_{\gamma}M$~\cite{Sakalli:2022xrb}. Therefore, as $M(t)$ evolves, a given value of $E_{\gamma}$ effectively ``samples'' a different region of the GBF, and therefore a different transmission probability. However, the functional form of the GBFs as a function of $E_{\gamma}M$ remains unchanged (so, loosely speaking, the GBFs do not ``evolve'').

We now consider a monochromatic population of primordial CCBHs, whose present-day comoving number density is $n(t_0)$. We compute the present-day intensity of photons emitted by the evaporation of these CCBHs across cosmic time, i.e.\ the rate per unit time per unit area per unit solid angle of photons with present-day energy $E_{\gamma 0}$, as follows:
\begin{align}
I(E_{\gamma 0})= &\frac{c}{4\pi}n(t_0)E_{\gamma 0} \nonumber \\
\times&\int^{z_{\star}}_{0} \frac{dz}{H(z)}\Phi_{\gamma}[(1+z)E_{\gamma 0},M(z)]\,,
\label{eq:intensity}
\end{align}
where $H(z)$ is the background expansion rate given by Eq.~(\ref{eq:hzlcdm}), and $z_{\star}$ is the redshift of recombination. Following the methodology adopted in Refs.~\cite{Calza:2024fzo,Calza:2024xdh,Calza:2025mwn} (as well as in several other works, including the seminal Ref.~\cite{Carr:2009jm}) we set upper limits on $n(t_0)$, which is the only unknown parameter in Eq.~(\ref{eq:intensity}), using measurements of the extragalactic $\gamma$-ray background (EGRB) from HEAO-1, COMPTEL, and EGRET in the $0.1 \lesssim E_{\gamma}/{\text{MeV}} \lesssim 100$ range~\cite{Gruber:1999yr,Schoenfelder:2000bu,Strong:2004ry}. The upper limit on $n(t_0)$ is the one for which the theoretically predicted intensity [Eq.~(\ref{eq:intensity})] first overshoots one of the measured EGRB datapoints. We then report constraints on the present-day CCBH density parameter $\Omega_{\text{ccbh}}=n(t_0)M(t_0)$. Unlike PBH-related works, we choose not to report constraints on $f_{\text{pbh}}$, the fraction of the DM in the form of PBHs, for two reasons. First, our primordial CCBHs are not necessarily a DM component, especially in light of our earlier discussions on the relation between $k$ and their effective equation of state. Next, these objects could in principle contribute to both the DM and DE component, so $f_{\text{pbh}}$ as usually computed would not be restricted to the range $f_{\text{pbh}}<1$, limiting the utility of this quantity. On the other hand, in a spatially flat FLRW Universe the present-day CCBH density parameter is only subject to the restriction $\Omega_{\text{ccbh}}<1-\Omega_b \approx 0.95$, given that the baryonic component contributes $\Omega_b \approx 0.05$ to the total energy budget. An important caveat is that, for sufficiently large values of $\Omega_{\text{ccbh}}$ approaching the theoretical upper limit of $1-\Omega_b$, our approximation of treating $H(z)$ as an externally specified cosmological background independently of the value of $k$ may not be entirely appropriate. This approximation is acceptable for the purposes of this part of our work, which does not aim to construct a fully self-consistent cosmological model of CCBHs, but rather to provide a first phenomenological estimate of evaporation constraints on the abundance of these objects, while leaving a complete self-consistent treatment to future work.~\footnote{We refer the reader to Refs.~\cite{Calza:2024fzo,Calza:2024xdh} for further discussions on a number of approximations and assumptions implicitly adopted in our analysis, their regime of validity, and why they are appropriate within the context of our work. These approximations include treating the distribution of CCBHs as being isotropic, considering only the primary photon spectrum, assuming a monochromatic mass distribution of CCBHs (see e.g.\ Refs.~\cite{Kuhnel:2015vtw,Kuhnel:2017pwq,Carr:2017jsz,Raidal:2017mfl,Bellomo:2017zsr,Lehmann:2018ejc,Carr:2018poi,Gow:2019pok,DeLuca:2020ioi,Gow:2020cou,Ashoorioon:2020hln,Bagui:2021dqi,Mukhopadhyay:2022jqc,Papanikolaou:2022chm,Cai:2023ptf}), as well as the methodology adopted for setting upper limits on $n(t_0)$ from EGRB observations.}

\section{Results and discussion}
\label{sec:discussion}

Let us first develop some useful intuition for the dynamics of the system governed by Eq.~(\ref{eq:dmdthrcc}), which reduces to Eq.~(\ref{eq:dmdthrcctgtcc}) for $t>t_{\text{CC}}$. Considering only $k>0$, the two effects clearly compete against one another: HR acts to decrease the BH mass, whereas the CC mechanism works to increase it. It is tempting to wonder whether there exists a ``balance mass'' for which the two effects balance each other out, leaving one with a BH whose mass is constant in time. In fact, the answer is negative, the reason being that the CC rate $\Gamma_{\text{CC}}$ explicitly depends on the expansion rate $H$, which evolves in time. It follows that a global, \textit{time-independent} balance mass does not exist.

Nevertheless, at any given time $\tilde{t}$, we can define an \textit{instantaneous balance mass} $M_{\star}(\tilde{t})$ by imposing $\dot{M}(\tilde{t}) \vert_{M_{\star}}=0$, with the dot indicating a time derivative. Using Eq.~(\ref{eq:dmdthrcctgtcc}), we find that $M_{\star}(t)$ is given by the following:
\begin{equation}
M_{\star}(t) = \left ( \frac{{\cal C}_{\text{HR}}}{kH(t)} \right ) ^{1/3}\,,
\label{eq:mstar}
\end{equation}
which is valid for $k>0$. At any given instant $\tilde{t}$, Eq.~(\ref{eq:mstar}) separates an HR-dominated evolution [$M(\tilde{t})<M_{\star}(\tilde{t})$] from a CC-dominated evolution [$M(\tilde{t})>M_{\star}(\tilde{t})$]. However, it is important to note that this is not a real equilibrium. Because $H(t)$ is evolving, so too does $M_{\star}(t)$ change with time. Therefore, a value of the BH mass which satisfies the instantaneous equilibrium condition at a given instant, will in general not satisfy this condition at earlier or later times. In the dynamical systems language, $M_{\star}(t)$ should therefore be understood as a moving nullcline, rather than a fixed point.~\footnote{We stress that only in the pure de Sitter case, where the Universe is dominated by a cosmological constant $\Lambda$ and the expansion rate $H=\sqrt{\Lambda/3}$ is a constant, does $M_{\star}=(3{\cal C}_{\text{HR}}^2/k^2\Lambda)^{1/6}$ truly become a constant, fixed point. For the $\Lambda$CDM Universe assumed here, this is the case only in the asymptotic future.}

In addition to being time-dependent, the instantaneous balance configuration associated with $M_{\star}(t)$ is also unstable. To see why, let us calculate $(\partial \dot{M}(M,t)/\partial M)_{M_{\star}}$, where $\dot{M}(M,t)$ is the right-hand side of Eq.~(\ref{eq:dmdthrcctgtcc}). Positive values of $(\partial \dot{M}(M,t)/\partial M)\vert_{M_{\star}}$ would imply that small departures from $M_{\star}$ are driven away from the instantaneous equilibrium point, and viceversa for negative values. Using Eqs.~(\ref{eq:dmdthrcctgtcc},\ref{eq:mstar}) we find the following (suppressing the time dependence of $M$ and $H$):
\begin{align}
& \left ( \frac{\partial \dot{M}}{\partial M} \right ) \Bigg\vert_{M=M_{\star}} = \left . \frac{\partial}{\partial M} \left ( -\frac{{\cal C}_{\text{HR}}}{M^2}+kHM \right ) \right \vert_{M=M_{\star}} \nonumber \\
={}& \left ( \frac{2{\cal C}_{\text{HR}}}{M^3}+kH \right ) \Bigg\vert_{M=M_{\star}} = 3kH\,,
\label{eq:instability}
\end{align}
which is always positive for $k>0$ and an expanding Universe. This implies that the instantaneous balance configuration defined by $M_{\star}(t)$ is unstable. Therefore, the ultimate outcome of the system must necessarily be either HR-dominated or CC-dominated, and no stable equilibrium configuration is possible. In the former case, the system evolves to $M(t) \to 0$ for $t \to t_{\text{ev}}$, whereas in the latter case it evolves to $M(t) \to \infty$ for $t \to \infty$. In principle, considering the asymptotic de Sitter future of our background cosmology, with \(H_{\infty}=\sqrt{\Lambda/3}\), there exists a critical separatrix trajectory such that $M \to M_{\star,\infty}=({\cal C}_{\text{HR}}/kH_{\infty})^{1/3}$ for $t \to \infty$, but it can be shown that this unstable and highly fine-tuned, and therefore not of physical interest.

\begin{figure*}[!htb]
\centering
\includegraphics[width=0.7\linewidth]{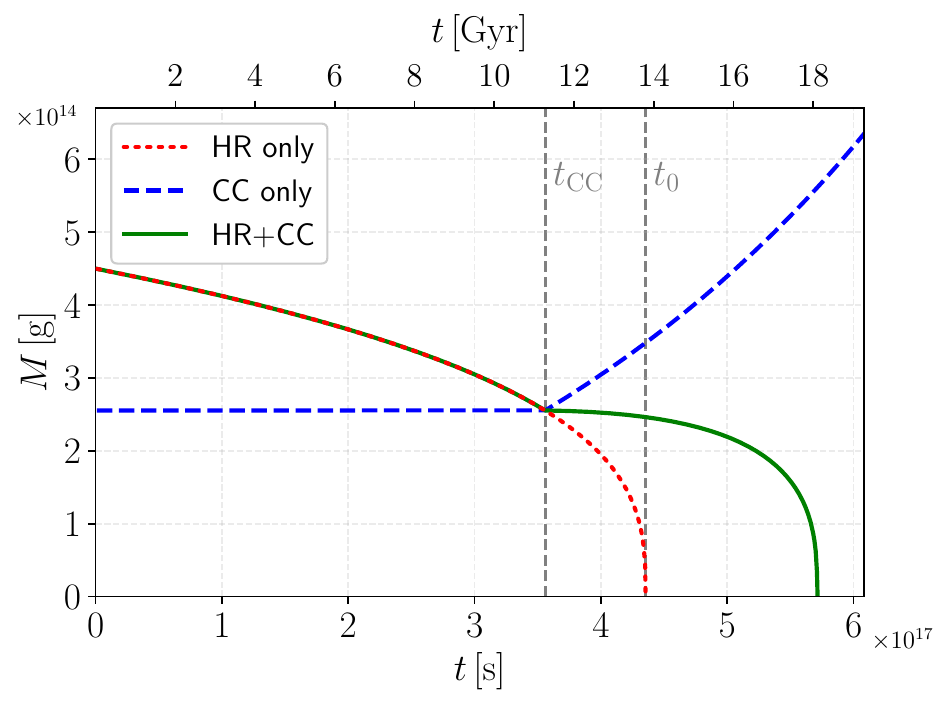}
\caption{Illustrative mass evolution of evaporating and/or cosmologically coupled BHs for three benchmark scenarios of interest, all of which are normalized to have the same value of mass at the time the CC mechanism activates, $M_{\text{CC}}=M(t_{\text{CC}})$, with the activation redshift being $z_{\text{CC}}=0.2$. The two vertical gray dashed lines indicate the time at which the CC mechanism activates (leftmost line, $t_{\text{CC}} \approx 11.5\,{\text{Gyr}}$) and the present time (rightmost line, $t_0 \approx 13.8\,{\text{Gyr}}$). The red dotted curve corresponds to a pure Hawking evaporation evolution of a BH whose mass at formation is $M_{\text{form}}=4.5\times 10^{14}\,{\text{g}}$: by $z_{\text{CC}}$, Hawking evaporation has reduced the BH mass to $M_{\text{CC}}\simeq 2.5\times 10^{14}\,{\text{g}}$, and the BH evaporates at $t_0$. The blue dashed curve instead corresponds to a pure CC evolution of a BH whose mass at formation is $M_{\text{form}}\simeq 2.5\times 10^{14}\,{\text{g}}$, where the coupling strength is $k=1.7$. Finally, the green solid curve shows the mass evolution of an evaporating cosmologically coupled BH, with the same formation mass as the red curve and the same CC parameters as the blue curve. For $t<t_{\text{CC}}$, the red and green curves overlap by construction. In this example the CC mechanism slows down the final stages of evaporation and prevents the BH from evaporating at $t_0$, prolonging its lifetime by $\approx 30\%$ to $t_{\text{ev}} \approx 18.0\,{\text{Gyr}}$.}
\label{fig:mevolution}
\end{figure*}

We now solve the mass evolution equation Eq.~(\ref{eq:dmdthrcc}) for a benchmark configuration of interest. Specifically, we first consider a BH with initial mass $M_{\text{form}}=4.5 \times 10^{14}\,{\text{g}}$: in the standard picture where the CC mechanism is absent, the lifetime of such a BH would be approximately the age of the Universe $t_0$, so the BH would be going through the final stages of Hawking evaporation right now. This pure HR evolution is the situation described by the red dotted curve in Fig.~\ref{fig:mevolution}, which indeed reaches $M \to 0$ as $t \to t_{\text{ev}}=t_0 \approx 13.8\,{\text{Gyr}}$. For concreteness and purely representative purposes, we then consider a CC activation redshift of $z_{\text{CC}}=0.2$, corresponding to an activation time of $t_{\text{CC}} \approx 11.5\,{\text{Gyr}}$, and a coupling strength $k=1.7$. By this time, a pure HR evolution has reduced the BH mass to $M_{\text{CC}} \approx 2.5 \times 10^{14}\,{\text{g}}<M_{\text{form}}$. We then choose to ``normalize'' the other curves in Fig.~\ref{fig:mevolution} to this value, in the sense that all three curves go through the same value of $M$ at $t_{\text{CC}}$. With this choice, the blue dashed curve corresponds to a pure CC-driven evolution, where HR is neglected (which is why the mass is constant for $t<t_{\text{CC}}$). Finally, the green solid curve corresponds to the case of interest to our work, i.e.\ evolution under the combined effects of HR and CC for a BH with with initial mass $M_{\text{form}}=4.5 \times 10^{14}\,{\text{g}}$, which would evaporate today in the absence of CC. For $t<t_{\text{CC}}$, being the CC mechanism not yet active, the red and green curves overlap by construction, since the evolution is driven only by HR. For $t>t_{\text{CC}}$ the CC mechanism starts competing against HR, and significantly slows down the mass loss: this is sufficient to prevent the BH from having fully evaporated at the present time (whereas, had the CC mechanism been absent, the BH would be reaching the endpoint of evaporation now), when its mass is $M_0 \approx M_{\text{CC}}$. For the specific configuration chosen, the ultimate outcome is still a HR-dominated one, but the BH lifetime increases by $\approx 30\%$, reaching $t_{\text{ev}} \approx 18.0\,{\text{Gyr}}$. While we have considered a specific benchmark, this simple example illustrates two important features:
\begin{itemize}
\item even a rather late activation of the CC mechanism (here occurring at $z_{\text{CC}}=0.2$) can drastically alter the fate of BHs which are otherwise extremely close to their evaporation threshold;
\item even when the ultimate outcome of the system is not CC-dominated, the BH lifetime can still be significantly increased by the CC mechanism.
\end{itemize}
As we shall see later, these features can have important phenomenological consequences.

\begin{figure*}[!htb]
\centering
\includegraphics[width=0.7\linewidth]{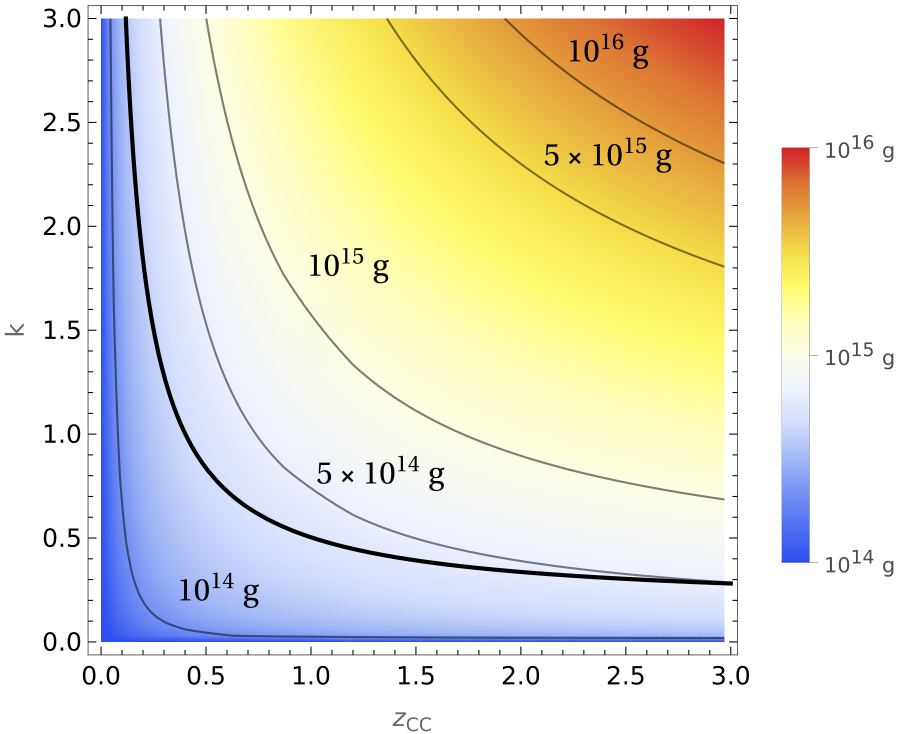}
\caption{Contour plot showing the present-day residual mass $M_0=M(t_0)$ of an evaporating cosmologically coupled BH whose formation mass is $M_{\text{form}}=4.5 \times 10^{14}\,{\text{g}}$, as a function of the CC activation redshift $z_{\text{CC}}$ (horizontal axis) and coupling strength $k$ (vertical axis). For this specific benchmark mass, an uncoupled BH would be evaporating at the present time: for this reason in the bottom-left corner, which corresponds to the uncoupled limit, the residual mass is $M_0=0$. Moving upward and/or to the right increases the effect of the CC mechanism, leading to larger residual masses today. The thick black curve corresponds to the locus where the present-day time derivative of the mass vanishes, i.e.\ the image in the $z_{\text{CC}}$-$k$ plane of the present-day instantaneous nullcline defined by Eq.~(\ref{eq:mstar}), $M(t_0)=M_{\star}(t_0)=({\cal C}_{\text{HR}}/kH_0)^{1/3}$. For points below this curve the present-day evolution is Hawking radiation-dominated ($\dot{M}(t_0)<0$), whereas for points above it the present-day evolution is CC-dominated ($\dot{M}(t_0)>0$). This curve is well approximated by the hyperbola $k(z_{\text{CC}}) \simeq 0.17+(3z_{\text{CC}})^{-1}$, and we stress that it should not be confused with the true separatrix between asymptotically Hawking radiation-dominated and CC-dominated trajectories. The contour lines connect points in the $z_{\text{CC}}$-$k$ plane with the same value of the present-day residual mass $M_0$.}
\label{fig:m0zk}
\end{figure*}

We now focus on the physical case described by the green curve, i.e.\ an evaporating CCBH with initial mass $M_{\text{form}}=4.5 \times 10^{14}\,{\text{g}}$. We generalize the example shown in Fig.~\ref{fig:mevolution} by scanning over the two parameters controlling the CC mechanism, i.e.\ the coupling strength $0 \leq k \leq 3$ and the activation redshift $0 \leq z_{\text{CC}} \leq 3$, showing the present-day mass $M_0$ in the contour plot of Fig.~\ref{fig:m0zk}. In a sense, Fig.~\ref{fig:m0zk} can therefore be viewed as the extension of the green curve of Fig.~\ref{fig:mevolution} to the whole $z_{\text{CC}}$-$k$ plane. The bottom left corner corresponds to the case with no CC, where the BH would be evaporating \textit{exactly} today, i.e.\ $M_0=0$. In the rest of the plot, the BH lifetime necessarily increases as discussed earlier, so $M_0>0$. We see that, for sufficiently large $k$ and/or $z_{\text{CC}}$, i.e.\ moving sufficiently up and/or to the right in Fig.~\ref{fig:m0zk}, the residual mass can be very large, up to $M_0 \sim {\cal O}(10^{16})\,{\text{g}}$ or larger. The fact that the residual mass $M_0$ can be larger than $M_{\text{form}}$ is not a problem, as this corresponds to cases where the present-day evolution is CC-dominated.

The black curve in Fig.~\ref{fig:m0zk} corresponds to the locus where the present-day time derivative of the mass vanishes, i.e.\ $\dot{M}(t_0)=0$. In other words, this curve is the image in the $z_{\text{CC}}$-$k$ plane of the present-day instantaneous nullcline defined by Eq.~(\ref{eq:mstar}), i.e.\ $M(t_0)=M_{\star}(t_0)=({\cal C}_{\text{HR}}/kH_0)^{1/3}$. Points below this curve correspond to BHs whose present-day evolution is HR-dominated [$\dot{M}(t_0)<0$], whereas points above it correspond to BHs whose present-day evolution is CC-dominated [$\dot{M}(t_0)>0$]. We stress that this curve does not correspond to the separatrix between asymptotically HR-dominated and CC-dominated trajectories, as it only captures the present-day sign of $\dot{M}$. Stated differently, a BH may lie above the black curve and hence be CC-dominated \textit{today}, but may still eventually enter the HR-dominated branch and evaporate at a finite future time. We find that the black curve is very well approximated by the hyperbola $k(z_{\text{CC}}) \simeq 0.17+(3z_{\text{CC}})^{-1}$. Finally, the contour lines in Fig.~\ref{fig:m0zk} connect points in the $z_{\text{CC}}$-$k$ plane which lead to the same present-day residual mass $M_0$.

\begin{figure*}[!htb]
\centering
\includegraphics[width=0.7\linewidth]{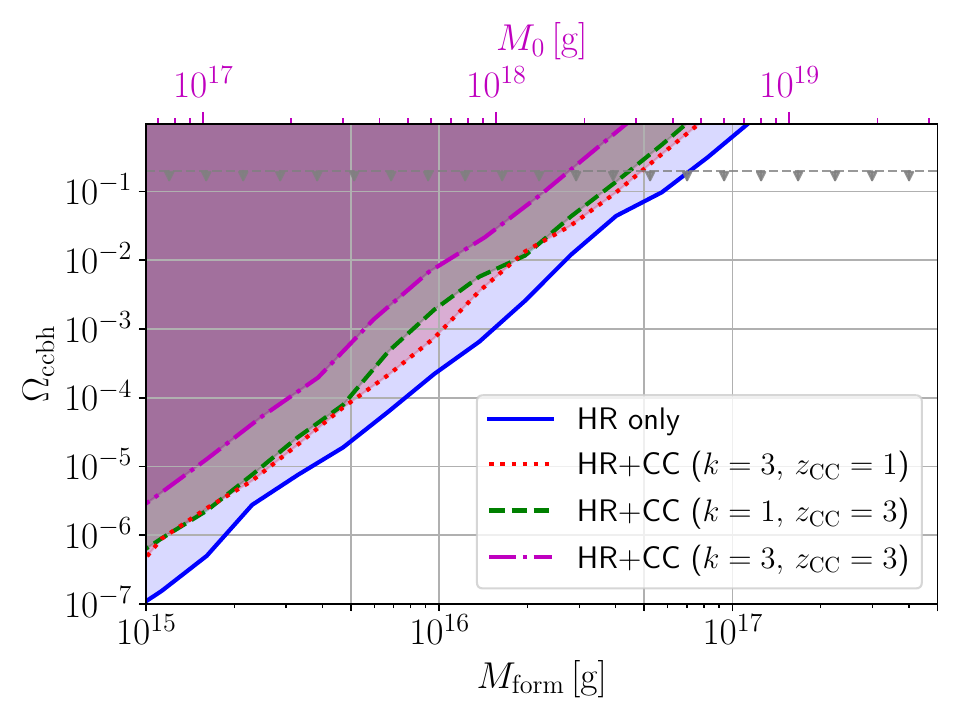}
\caption{Upper limits on $\Omega_{\text{ccbh}}$, the present-day CCBH density parameter, as a function of the CCBH formation mass $M_{\text{form}}$. The limits are derived for different evaporating CCBH benchmarks: $k=3$ and $z_{\text{CC}}=1$ (red dotted curve), $k=1$ and $z_{\text{CC}}=3$ (green dashed curve), and $k=3$ and $z_{\text{CC}}=3$ (magenta dash-dotted curve). Note that the blue solid curve corresponds to the standard case where only Hawking radiation is at play, which is recovered in the limit where $k=0$ and $z_{\text{CC}}=0$. The upper horizontal axis shows the present-day CCBH mass $M_0$, computed for the specific $k=3$ and $z_{\text{CC}}=3$ evaporating CCBH benchmark, and is colored in magenta to reinforce the connection to the magenta benchmark. This upper horizontal axis is plotted simply for representative purposes, and is useful for assessing whether limits depending on the present-day mass of the CCBH population, such as microlensing limits, may also become relevant. The gray dashed horizontal line with downward-pointing arrows corresponds to $\Omega_{\text{ccbh}}=0.2$, and is intended as a conservative visual guide marking the region (above the line) where the fixed background approximation should be interpreted with more caution: below the line (gray arrows pointing downwards), such an approximation is instead considerably safer.}
\label{fig:omegaccbh}
\end{figure*}

Finally, in Fig.~\ref{fig:omegaccbh} we show upper limits on the present-day (primordial) CCBH density parameter $\Omega_{\text{ccbh}}$, as a function of both the primordial CCBH formation mass $M_{\text{form}}$ (lower horizontal axis) and present-day mass $M_0$ (upper horizontal axis), the latter computed for the specific $k=3$ and $z_{\text{CC}}=3$ evaporating CCBH benchmark. The blue solid curve corresponds to the standard case where only HR is present, and is the baseline against which we compare the effects of the CC mechanism. For simplicity, we consider three evaporating CCBH benchmarks, characterized by the following values of the CC parameters: $k=3$ and $z_{\text{CC}}=1$ (red dotted curve), $k=1$ and $z_{\text{CC}}=3$ (green dashed curve), and $k=3$ and $z_{\text{CC}}=3$ (magenta dash-dotted curve). The gray dashed horizontal line with downward-pointing arrows marks $\Omega_{\text{ccbh}}=0.2$, and is intended as a conservative visual guide marking the region (above the line) where the fixed background approximation should be interpreted with more caution. Below the line, such an approximation is instead considerably safer. We stress that this value is not meant to define a sharp validity threshold, but is just intended as a conservative and approximate visual guide. For substantially larger abundances, the fixed $H(z)$ approximation should be interpreted with increasing caution. As expected from the previous discussion, turning on the CC mechanism necessarily weakens the evaporation constraints. In other words, at fixed $M_{\text{form}}$, the allowed abundance of CCBHs is larger relative to the uncoupled case. The reason is that the CC mechanism keeps the CCBH far from the (runaway) endpoint of the evaporation regime for a longer time, thereby reducing the intensity of the photons emitted during the evaporation. We stress that this intensity, and therefore the EGRB signal, is sensitive to the entire mass history $M(z)$ entering Eq.~(\ref{eq:intensity}), rather than just the formation mass or the present-day mass.

A few points concerning the above limits and their physical interpretation are worthy of further discussion. We first note that, in the usual studies of PBH evaporation constraints, for most of the parameter space considered here one typically neglects the difference between $M_{\text{form}}$ and $M_0$, and treats the evaporating PBHs as being quasi-static over the age of the Universe, at a single mass $M_{\text{pbh}}$. In other words, in the integral of Eq.~(\ref{eq:intensity}) one typically sets $M(z)=M_{\text{pbh}} \simeq M_{\text{form}} \simeq M_0$. This is a good approximation because Hawking evaporation is a runaway process, where most of the BH mass is lost in the very final instants of evaporation. For $M_{\text{pbh}} \gtrsim 10^{15}\,{\text{g}}$, an uncoupled evaporating BH has a lifetime of $t_{\text{ev}} \sim 10t_0$, and reaches $t_0$ with $M_0 \gtrsim 0.95M_{\text{form}}$. This approximation is of course not generally valid in the case of evaporating CCBHs, and in Eq.~(\ref{eq:intensity}) we thus track the full evolution of $M(z)$ throughout the redshift integral. For this reason, our limits should not be interpreted as a simple ``rescaling'' of the usual PBH evaporation limits, given that the CC mechanism can significantly change the relation between the formation mass $M_{\text{form}}$, the mass controlling the emission at a given redshift $M(z)$, and the present-day mass $M_0$.

This brings us to a related important point. It is important to distinguish between evaporation limits, which are sensitive to the entire CCBH emission history (and therefore its entire mass evolution throughout cosmic time), and other limits depending primarily on the PBH mass at late times, such as microlensing limits, which essentially depend on $M_0$. If the CC mechanism is sufficiently strong, i.e.\ if $k$ and/or $z_{\text{CC}}$ are sufficiently large, an evaporating CCBH with $M_{\text{form}}$ in the range usually associated with evaporation constraints may have a present-day mass $M_0$ for which microlensing constraints become relevant. In other words, the same CCBH population, or even the same point in $(M_{\text{form}},k,z_{\text{CC}})$ parameter space, may in principle be \textit{simultaneously} subject to evaporation limits because of its emission history and microlensing limits because of its present-day mass: whether this is the case can be checked from the upper horizontal axis in Fig.~\ref{fig:omegaccbh} (which, we stress, should be recomputed for each CCBH benchmark, as it explicitly depends on both $k$ and $z_{\text{CC}}$). For the same reason, the point on the lower horizontal axis where the evaporation limit on $\Omega_{\text{ccbh}}$ reaches the present-day DM density parameter $\Omega_{\text{dm}}$ should not be interpreted as the lower edge of the asteroid mass window. The usual notion of asteroid mass window is defined for standard PBHs behaving as a pressureless DM component, for which different constraints can be mapped onto a single mass parameter $M_{\text{pbh}}$. This one-to-one mapping no longer holds for CCBHs, so limits which normally apply to different regions of $M_{\text{pbh}}$ parameter space, such as evaporation vs microlensing limits, may now simultaneously apply to a given CCBH population, even if this is monochromatic. A complete investigation of non-evaporation constraints on the full CCBH parameter space requires a dedicated study which is beyond the scope of this paper, and is therefore left to future work.

\section{Conclusions}
\label{sec:conclusions}

Recent observational and theoretical advances have led to cosmologically coupled black holes (CCBHs) receiving tremendous interest, as these objects may play a role in addressing both the singularity and cosmic acceleration problems, while leading to a host of other unique observational signatures~\cite{Faraoni:2007es,Faraoni:2007gq,Faraoni:2008tx,Davidson:2012si,Croker:2019mup,Croker:2019kje,Croker:2020plg,Croker:2021duf,Cadoni:2022chn,Farrah:2022ghw,Farrah:2023opk,Mistele:2023fds,Cadoni:2023lum,Rodriguez:2023gaa,Parnovsky:2023wkc,Avelino:2023rac,Wang:2023aqe,Andrae:2023wge,Lei:2023mke,Sadeghi:2023cpd,Ghodla:2023iaz,Garcia-Bellido:2024tip,Yagdjian:2023yjf,Amendola:2023ays,Gao:2023keg,Gaur:2023hmk,Cadoni:2023lqe,Mlinar:2023fkk,Lacy:2023kbb,Dahal:2023hzo,Croker:2024jfg,Kovacik:2024xva,Cadoni:2024jxy,Lu:2024ppa,Faraoni:2024ghi,Calza:2024qxn,DES:2024ffp,Cadoni:2024rri,Farrah:2025ghw,Ahlen:2025owq,DESI:2025ffm,Lei:2025qff,Hayashi:2025frr,deLima:2025bqf,FelipeReis:2025qpf,Cadoni:2026ejk,Abilmazhinova:2026cog,Wu:2026xwo,Milos:2026ayt,Zhou:2026dti}. However, all existing studies on CCBHs treat these as purely classical objects, neglecting Hawking radiation (HR). Evaporation can have a significant impact on the evolution of CCBHs, as it acts to decrease the CCBH mass, thereby competing with the cosmological coupling (CC) mechanism, which instead acts to increase the CCBH mass. Our goal in this work has therefore been to take a first step towards studying evaporating CCBHs and their phenomenology. We have argued that the problem is most naturally tackled within a quasi-adiabatic approximation. Specifically, we evolve the CCBH mass under the combined effects of CC (via the usual power-law mass growth ansatz) and HR, with the instantaneous evaporation rate taken to be the standard Hawking one evaluated at the corresponding value of the evolving CCBH mass. The validity of this approximation does not require a hierarchy between the CC and HR rates, but rather that the mass evolution be slow compared to the typical CCBH dynamical timescale, in which case the CCBH can be treated as evolving through a sequence of quasi-static configurations.

We have shown that the competition between HR and CC cannot lead to a stable balance configuration: the system is instead characterized by a time-dependent instantaneous balance mass, which defines an unstable configuration. The asymptotic outcome of the evolution of evaporating CCBHs is therefore either CC-dominated or HR-dominated. In the former case, the CCBH mass is ever-increasing as time goes on. In the latter case, the CCBH ultimately evaporates at a finite time, but its lifetime is increased, relative to the uncoupled case, by the CC mechanism. In this case we have shown that even a CC mechanism of modest strength, or activating at rather late times, can significantly extend the BH lifetime: for instance, the lifetime of a primordial BH whose mass at formation is $M=4.5 \times 10^{14}\,{\text{g}}$, and which ordinarily would be evaporating at the present time~\cite{Khlopov:2008qy,Carr:2020xqk,Green:2020jor,Choudhury:2024aji,Shankaranarayanan:2026hnn}, can increase by $30\%$ for a coupling strength of $k=1.7$ and a very late CC activation at redshift $z_{\text{CC}}=0.2$ (see Fig.~\ref{fig:mevolution}). Finally, we have computed limits on the abundance of primordial, evaporating CCBHs (see Fig.~\ref{fig:omegaccbh}): these are weaker relative to their uncoupled counterparts, since the CC mechanism keeps the BH far from the (runway) endpoint of the evaporation regime for a longer time, reducing the intensity of the emitted photons. Finally, we have argued that the interpretation of evaporation limits can be significantly more subtle than in the uncoupled case, given that the formation mass and present-day mass no longer coincide (unlike in the uncoupled case, where they approximately coincide, at least for $M_{\text{pbh}} \gtrsim 10^{15}\,{\text{g}}$). As a result of this, an evaporating CCBH population subject to evaporation limits because of its emission history may simultaneously be subject to other constraints, such as microlensing ones, if the CC mechanism is sufficiently strong and hence the present-day mass sufficiently large. We also recall that the interested reader can find a number of useful analytical expressions, including a closed-form analytical solution for the mass evolution of evaporating CCBHs, in Appendix~\ref{app:analytical}.

We stress that our work should be viewed as a first exploration of evaporating CCBHs, whose phenomenology, as already shown by our results, is rather rich. Several open questions and interesting directions for future research remain, some of which we mention here. Firstly, we were forced to adopt our quasi-adiabatic approximation given the lack of a fully dynamical space-time describing an evaporating cosmologically coupled BH. Theoretical studies in this direction are thus very much in need, even in the simpler uncoupled case, where one typically resorts to phenomenological ans\"{a}tze for the mass growth, rather than studying a first-principles CCBH metric. At the same time, much more work is required to clarify the role and impact of spin on the CC mechanism, including the question of whether spin itself is cosmologically coupled: the answers to these questions are at present completely unknown, but can play a key role in the study of evaporating CCBHs, given the significant impact of spin on the evaporation of uncoupled BHs. Concerning primordial CCBHs, while we have only studied evaporation limits thereon, an important next step would be to systematically revisit other sources of limits usually studied for Schwarzschild PBHs, such as microlensing, accretion, and dynamical limits~\cite{Khlopov:2008qy,Carr:2020xqk,Green:2020jor,Choudhury:2024aji,Shankaranarayanan:2026hnn}. Because of the CC mechanism, more than one of these limits may apply at the same time for a given formation mass, depending on the values of $k$ and/or $z_{\text{CC}}$, so the question of what is the most efficient way of presenting these limits becomes a very relevant one. Moreover, in determining the expansion history $H(z)$ we have essentially adopted a ``test population'' approximation, where the CCBHs are not the main source of the expansion. When $\Omega_{\text{ccbh}}$ approaches $1-\Omega_b$ this approximation clearly ceases to be valid: in this case, the expansion history should self-consistently account for the CC mechanism, and would depend on $k$. Finally, it might be worth exploring effects which are potentially relevant in the latest stages of the evaporation of CCBHs, such as the memory burden effect~\cite{Dvali:2020wft,Thoss:2024hsr,Kohri:2024qpd,Zantedeschi:2024ram,Chianese:2024rsn,Montefalcone:2025akm,Dvali:2025ktz,Chianese:2025wrk,Dondarini:2025ktz,Dvali:2026tia}, as this can have a significant impact on the results presented here. We leave these and related interesting research directions to future work.

\begin{acknowledgments}
\noindent We acknowledge support from the Istituto Nazionale di Fisica Nucleare (INFN) through the Commissione Scientifica Nazionale 4 (CSN4) Iniziativa Specifica ``Quantum Fields in Gravity, Cosmology and Black Holes'' (FLAG). This publication is based upon work from the COST Action CA21136 ``Addressing observational tensions in cosmology with systematics and fundamental physics'' (CosmoVerse), supported by COST (European Cooperation in Science and Technology).
\end{acknowledgments}

\appendix

\section{Analytical expressions}
\label{app:analytical}

Here we show that, within the flat $\Lambda$CDM background assumed in this work, the mass evolution equation for evaporating CCBHs after the CC mechanism activates [Eq.~(\ref{eq:dmdthrcctgtcc})] admits a closed-form analytical solution. The starting point is the cosmic time-redshift relation given by Eq.~(\ref{eq:tz}). Within the late-time approximation to the expansion rate given by Eq.~(\ref{eq:hzlcdm}), where the radiation component is neglected, the integral admits a well-known analytical solution, given by the following:
\begin{equation}
t(z)=\varpi^{-1}\operatorname{arcsinh} \left [ \sqrt{\frac{1-\Omega_m}{\Omega_m}}(1+z)^{-\frac{3}{2}} \right ] \,,
\label{eq:tzlcdm}
\end{equation}
where the $(1-\Omega_m)$ factor appears since the cosmological constant density parameter is $\Omega_{\Lambda} \approx (1-\Omega_m)$ within the approximation where radiation is neglected, and the quantity $\varpi$ is defined as follows:
\begin{equation}
\varpi \equiv \frac{3}{2}H_0\sqrt{1-\Omega_m}\,.
\label{eq:varpi}
\end{equation}
To solve the mass evolution equation, Eq.~(\ref{eq:dmdthrcctgtcc}), we invert Eq.~(\ref{eq:tzlcdm}) and obtain the following:
\begin{equation}
z(t)= \left ( \frac{1-\Omega_m}{\Omega_m} \right ) ^{\frac{1}{3}} \sinh^{-\frac{2}{3}}\left ( \varpi t \right ) -1\,.
\label{eq:ztlcdm}
\end{equation}
Recalling that the Hubble rate is defined as $H \equiv \dot{a}/a$, with the dot denoting time derivatives, we note that Eq.~(\ref{eq:ztlcdm}) can be used to obtain the Hubble rate as a function of time as follows:
\begin{align}
H(t) &= -\frac{\dot{z}}{1+z} = \frac{2}{3}\varpi \coth \left ( \varpi t \right ) \nonumber \\
&= H_0\sqrt{1-\Omega_m}\coth \left ( \varpi t \right ) \,.
\label{eq:htlcdm}
\end{align}
The negative sign in the first equality accounts for the fact that, as time progresses, $z$ decreases, so that $H(t)$ is a positive quantity.

We now solve Eq.~(\ref{eq:dmdthrcctgtcc}) setting boundary conditions at the CC activation time, i.e.\ $M(t_{\text{CC}})=M_{\text{CC}}$. The equation can be cast into a simple form using the substitution $Y(t)\equiv M^3(t)$ and multiplying by $3M^2$, from which the following linear equation for $Y(t)$ is obtained:
\begin{equation}
\frac{dY}{dt}-3kH(t)Y=-3{\cal C}_{\text{HR}}\,.
\label{eq:y}
\end{equation}
Using the standard integration factor method, and the fact that $a=(1+z)^{-1}$, Eq.~(\ref{eq:y}) can be solved to obtain the following:
\begin{align}
Y(t)= {}& \left ( \frac{1+z_{\text{CC}}}{1+z(t)} \right ) ^{3k} \nonumber \\
\times {}& \left [ M_{\text{CC}}^3-3{\cal C}_{\text{HR}}\int_{t_{\text{CC}}}^{t}dt' \left ( \frac{a(t')}{a(t_{\text{CC}})} \right ) ^{-3k} \right ] \,,
\label{eq:ysolution}
\end{align}
from which we can directly read off the solution for $M(t)$:
\begin{equation}
M(t)=\left ( \frac{1+z_{\text{CC}}}{1+z(t)} \right ) ^k
\left [ M_{\text{CC}}^3-\frac{3{\cal C}_{\text{HR}}}{(1+z_{\text{CC}})^{3k}}
{\cal I}_k(t) \right ] ^{\frac{1}{3}}\,,
\label{eq:msolution}
\end{equation}
with the integral ${\cal I}_k(t)$ defined as follows:
\begin{equation}
{\cal I}_k(t)\equiv \int_{t_{\text{CC}}}^{t}dt'\, [1+z(t')]^{3k}\,.
\label{eq:ik}
\end{equation}
Using Eq.~(\ref{eq:ztlcdm}), this integral can be expressed analytically as follows:
\begin{equation}
{\cal I}_k(t) = \left ( \frac{1-\Omega_m}{\Omega_m} \right ) ^k \frac{\varsigma_k(t)-\varsigma_k(t_{\text{CC}})}{\varpi}\,,
\label{eq:ikanalytical}
\end{equation}
where the function $\varsigma_k(t)$ is defined as follows:
\begin{align}
\varsigma_k(t)={}& \frac{\sinh^{1-2k}\left ( \varpi t \right ) }{1-2k} \nonumber \\
\times {}&_2F_1\left ( \frac{1}{2},\frac{1}{2}-k;\frac{3}{2}-k;-\sinh^2\left ( \varpi t \right ) \right ) \,,
\label{eq:varsigma}
\end{align}
and $_2F_1$ is the hypergeometric function. We have checked that the analytical solution given by Eq.~(\ref{eq:msolution}) is in excellent agreement with our numerical solution. We also note that, in the integrals appearing in Eqs.~(\ref{eq:ysolution},\ref{eq:ik}), we have fixed the lower limits of integration to $t_{\text{CC}}$. This essentially fixes the integration constants which would otherwise appear in the general indefinite integral solution, and guarantees that the boundary condition $Y(t_{\text{CC}})=M_{\text{CC}}^3$ is automatically satisfied.

As a further sanity check, we have also verified that we can recover the well-known pure-CC and pure-HR solutions for $M(t)$ in the appropriate limits of Eq.~(\ref{eq:msolution}). In the pure-CC case, we set ${\cal C}_{\text{HR}}=0$ in Eq.~(\ref{eq:msolution}), from which we immediately obtain the following:
\begin{equation}
M(t)=M_{\text{CC}}\left ( \frac{1+z_{\text{CC}}}{1+z(t)} \right ) ^k\,.
\label{eq:msolutionpurecc}
\end{equation}
which matches the CC mass growth ansatz adopted in this work [Eq.~(\ref{eq:mccz})]. On the other hand, the pure-HR case is obtained setting $k=0$ in Eq.~(\ref{eq:msolution}). In this case, the ${\cal I}$ integral trivially reduces to ${\cal I}_0(t)=t-t_{\text{CC}}$ which, once inserted into Eq.~(\ref{eq:msolution}), gives the following:
\begin{equation}
M(t)=\left [ M_{\text{CC}}^3-3{\cal C}_{\text{HR}}(t-t_{\text{CC}}) \right ] ^{\frac{1}{3}}\,.
\label{eq:msolutionpurehr}
\end{equation}
which matches the pure-HR solution [Eq.~(\ref{eq:mhrt})]. We stress that the analytical solution given in Eq.~(\ref{eq:msolution}) is valid only for the late-time approximation to the spatially flat $\Lambda$CDM expansion rate adopted throughout our work, and does not easily generalize to other cosmologies. Nevertheless, we have chosen to report it as it may be of interest in its own right.

\clearpage
\bibliography{hrccbh}

\end{document}